\documentstyle[12pt]{article}  
  
\def\a{\alpha} 
\def\b{\beta}  
 
\def\s{\sigma}  
\def\d{\delta}

\def\m{\mu}  
\def\n{\nu}

\def\g{\gamma}  
\font\mybb=msbm10 at 12pt  
\def\bb#1{\hbox{\mybb#1}}  
\def\bZ {\bb{Z}} 
\def\bR {\bb{R}}

\def\s9{{\scriptscriptstyle{9}}}
\def\h10{{\widehat{10}}}
\def\hA{{\hat A}}
\def\hG{{\hat G}}

\newcommand{\be}{\begin{equation}} 
\newcommand{\ee}{ \end{equation}}
\newcommand{\ba}{\begin{array}}
\newcommand{\ea}{\end{array}}
\newcommand{\bea}{\begin{eqnarray}}
\newcommand{\eea}{\end{eqnarray}}
\newcommand{\ft}[2]{{\textstyle\frac{#1}{#2}}}
\newcommand{\eqn}[1]{(\ref{#1})}
\renewcommand{\a}{\alpha}
\renewcommand{\b}{\beta}

\renewcommand{\d}{\delta}
\newcommand{\pa}{\partial}
\newcommand{\G}{\Gamma}

\begin{document}  
  
\begin{titlepage}

\begin{flushright}  
AEI-1999-13\\  
HUB-EP-99/43\\
THU-99/22\\[1mm] 
{\tt hep-th/9908169}  
\end{flushright}  
  
\vspace{1cm}  
\begin{center}  
  
{\Large \bf{Space-Time Supersymmetry, IIA/B Duality and M-Theory}}  
  
\vspace{1.6cm}  
  
Mohab Abou-Zeid$^1$\footnote{abouzeid@aei-potsdam.mpg.de}, 
Bernard de Wit$^{1,2\,}$\footnote{bdewit@phys.uu.nl}, 
Dieter L\"{u}st$^3$\footnote{luest@physik.hu-berlin.de} and 
Hermann Nicolai$^1$\footnote{nicolai@aei-potsdam.mpg.de}  
  
\vspace{.6cm}  
  
$^1${\em Max Planck Institut f\"{u}r Gravitationsphysik, Albert
Einstein Institut,\\Am M\"{u}hlenberg 5, D-14476 Golm, Germany}  
  
\vspace{.3cm}  
$^2${\em Institute for Theoretical Physics, Utrecht University,\\
Princetonplein 5, 3508 TA Utrecht, The Netherlands} 

\vspace{.3cm}  
$^3${\em Institut f\"{u}r Physik, Humboldt Universit\"{a}t zu Berlin,
\\ Invalidenstrasse, D-10115   Berlin, Germany}   

\vspace{.6cm}  
\begin{abstract}
The connection between IIA superstring theory compactified on a circle
of radius $R$ and IIB theory compactified on a circle of radius
$1/R$ is reexamined from the perspective of $N=2$, $D=9$ space-time
supersymmetry. We argue that the consistency of  IIA/B duality
requires the BPS states corresponding to momentum and winding of
either of the type-II superstrings  to transform as inequivalent
supermultiplets. We show that this is indeed the case for  any
finite compactification radius, thus providing a nontrivial
confirmation of IIA/B duality. From the point of view of $N=2$, $D=9$
supergravity, one is naturally led to an SL$(2,\bZ)$ invariant field
theory that encompasses both the M-theory torus and the Kaluza-Klein
states of the IIB theory.   
\end{abstract}
  
\end{center}  
\vspace{.5cm}
 
\flushleft{August, 1999}
 
\end{titlepage}  
  
The bosonic string compactified
on a circle of radius $R$ is subject to a duality which relates the
theory obtained at a compactification radius $R$ to the theory
compactified at radius $1/R$, where we set the string scale to
unity~\cite{KYSS}.  The origin of this duality is that 
momentum modes, whose masses are  multiples of $1/R$ are
accompanied by winding modes, whose masses are  multiples
of $R$. The spectrum exhibits a  symmetry under
$R\to 1/R$ combined with an interchange of momentum and winding
states, which in fact is a symmetry of the full string
theory. Because large and small distances are related, $R$ can be
restricted to the interval $[1,\infty)$ and the theory appears to have
a smallest length set by the string scale. At $R=\infty$ the theory
decompactifies while at the self-dual point $R=1$ the 
winding and momentum states acquire equal masses and gauge symmetry
enhancement takes place. The heterotic string possesses the same kind
of duality symmetry~\cite{Gins}.

However, there are situations where string theory is not self-dual in
this naive sense, although the spectra at compactification scales $R$
and $1/R$ seem to be identical. This is the case for  the type-II string theories~\cite{DHS,DLP}. One way to
analyze whether or not the theory is   
self-dual is to start at large compactification
radius and to extrapolate all the way to zero radius. In that limit,
the winding states become massless and the theory is expected to again 
decompactify. If this is indeed the case, one must obtain one of the
consistent string theories defined in the uncompactified
space-time; either this is the theory one started with, or
it is a different string theory. 

The approach followed in this paper 
is that one can understand which theories are related by duality
without extrapolating  to zero radius provided the
winding and momentum states carry 
different space-time quantum numbers from the very beginning, where by
space-time we refer to the generic space-time with one compactified
coordinate of finite radius. We claim that this happens to the
type-II string 
theories, where we will show that the momentum and winding states
constitute {\em inequivalent} $N=2$, $D=9$ supermultiplets at any
given compactification 
radius. In this situation two immediate conclusions are obvious. First
of all, there will be no symmetry enhancement when 
the masses of momentum and winding states become equal (this is
consistent with the fact that the underlying conformal theory does
not give rise to gauge symmetry enhancement), and secondly,
the uncompactified theory obtained at $R=0$ is distinct from the
theory at $R=\infty$. In this case,  duality is conceptually different from a
symmetry. Clearly one must find a different theory for any value of
the compactification radius. There are no two radii at which the
corresponding theories
could conceivably be identical, because their spectra 
will be inequivalent; thus the different theories are now parametrized by
the radius $R$ in the interval $(0,\infty)$. Duality  means that 
two theories that are unrelated in the uncompactified space-time,
can be viewed as different limits in the \lq moduli' space of the 
compactified theories. Of
course, one can  describe the compactified theory
from the perspective of one of the two inequivalent uncompacified theories
associated with the endpoints $R=\infty$ and $R=0$, but this leaves
it unchanged. 

The present evidence for IIA/B duality is either based on formally
mapping one theory onto the other, or on studying the behaviour near
the two decompactification points at $R=\infty$ and
$R=0$. In~\cite{DHS} it was argued that the consistency of the 
interchange of momentum and winding numbers, and of $R$ and $1/R$, with
world-sheet superconformal invariance requires that the 
components in the compactified direction of both the left-moving
bosonic and fermionic world-sheet fields change sign, leading to a
corresponding flip in the GSO projection~\cite{GSO}.
Formally, one thus obtains a mapping between two consistent theories,
namely compactified IIA and compactified IIB string theory, which extends
to their respective vertex operators. In \cite{DLP} a possible
continuous connection between the 
two theories is investigated by considering the Lorentz generators
near the two decompactification points. The sign change of the
world-sheet fields is  invoked in order to show that the
Lorentz  representations carried by the Ramond-Ramond ground state in these two
limits must be different. In this approach, however (as  noted
in \cite{DLP}),  the ten-dimensional Lorentz invariance is
broken when the compactification radius $R$ is different from zero or
infinity, so that one cannot truly interpolate between the two theories.

The arguments given in this paper provide additional  evidence
that the IIA and the IIB 
theories are asymptotic limits in a one-parameter moduli space of
theories. First, we analyze the nine-dimensional $N=2$
supersymmetry algebra from the space-time viewpoint and  on the basis
of perturbative string theory and M-theory. This analysis 
indeed confirms that the winding and momentum states constitute
inequivalent supermultiplets in  either of the type-II string theories
at a given 
compactification radius. We then show that the same conclusion follows
upon imposing the physical state conditions on the relevant string
vertex operators. In fact,  $N=2$, $D=9$ supersymmetry alone already
gives rise to a unified description encompassing 
both the M-theory torus and the Kaluza-Klein states
of the IIB theory. Our work lends support to the arguments
given in \cite{JHS} that there is a duality between M-theory and IIB
theory which can be understood  in terms of the fundamental supermembrane
\cite{BST}. The coupling of one class of BPS states to 
supergravity breaks the continuous SL$(2,\bR)$ symmetry group into
a discrete subgroup associated with the Kaluza-Klein states on
$T^2$. The second class of BPS multiplets carries charges that are
unrelated  to $T^2$ and can be identified with either the Kaluza-Klein
states of the IIB theory, or the wrapping of a membrane around the
torus.

\vspace{.5cm} 

\noindent{\it N=2 Supersymmetry in  Nine Dimensions}\\  
Let us first summarize the various possible BPS multiplets associated
with the nine-dimensional supersymmetry algebra with Lorentz invariant
central charges. We consider the
supersymmetry algebra in an $N=2$ Majorana basis and write it as follows, 
\be
\{Q_\a^i\,, Q_\b^{\dagger \,j} \} =  (P\!\llap / \,\g^0)_{\a\b} +
Z^{ij}\, (i\g^0)_{\a\b} \, .
\label{susy}
\ee
We recall that, in nine dimensions,  the
charge conjugation matrix is symmetric and can be chosen equal to the
identity; therefore $Z^{ij}$ is a real {\it symmetric} matrix. We can
decompose the central charge as
\be 
Z^{ij} = M\,[\,b \,\delta^{ij} +  a\, (\cos \theta \,\sigma_3 + \sin
\theta \,\sigma_1)^{ij}], \label{Zmatrix}
\ee
where $M$ is the rest mass of the
representation and $\sigma_1 ,\sigma_2 ,\sigma_3$ denote the usual Pauli matrices. It is clear, and this is crucial for what  
follows, that these central charges fall into two categories. The
component proportional to $b$ defines an SO(2) invariant central
charge, while the two components proportional to $a$ rotate into each
other under the action of the automorphism group SO(2). 

Let us first derive the possible values for $Z^{ij}$ when
straightforwardly reducing the ten-dimensional supersymmetry
algebra. We first decompose the Clifford algebra 
generated by the ten-dimensional gamma matrices $\Gamma^M$, and define 
$\Gamma^{11}$ by 
$\Gamma^{11} = \Gamma^{0}\Gamma^{1}\cdots
\Gamma^{9}$. Nine-dimensional gamma matrices, which commute with 
$\Gamma^9$ and $\Gamma^{11}$, are given by
\be
\gamma^\m = \G^\m \tilde \gamma\,,\qquad \m =0, 1, 2, \ldots,8,
\ee
where $\tilde\gamma = -i\G_9 \G_{11}$, such that the product
$\g^0\g^1\cdots\g^8 = -i {\bf 1}$. Observe that the Dirac conjugate of
a spinor is changed accordingly.\footnote{
In nine and ten dimensions, the Dirac conjugate is defined by $\bar
\psi= i\psi^\dagger \gamma^0$ and $\bar \psi= i\psi^\dagger
\Gamma^0$, respectively. Note that the $\gamma^\m$ and
$\Gamma^M$ are hermitean, with the exception of $\gamma^0$ and
$\Gamma^0$, which are 
anti-hermitean. We use the `mostly plus' metric. Observe that our
conventions are such that there is no  $\G^{10}$ matrix. 
} 
The ten-dimensional charge-conjugation
matrix can now be written as $\tilde \gamma$, so that the
nine-dimensional gamma matrices are symmetric and the nine-dimensional
charge-conjugation matrix equals the unit matrix.
With these conventions,
the ten-dimensional supersymmetry algebra $\{Q,\bar Q\}= -i P_M
\Gamma^M$ is converted into the nine-dimensional algebra $\{Q,\bar
Q\}= -i P_\m \gamma^\m - P_9 \,\Gamma^{11}$. Hence the sign of the
$P_9$ term depends on the chirality of the supercharge. Therefore, the
matrix $Z^{ij}$ will be proportional to $(\sigma_3)^{ij}$ for IIA theory,
where one has supercharges of opposite chirality, and proportional to
$\delta^{ij}$ for IIB theory, where the charges have equal chirality. 

To exhibit the BPS multiplets in nine dimensions we diagonalize 
the matrix \eqn{Zmatrix} by an appropriate SO(2) transformation 
so that $\theta=0$. In the rest frame the anticommutator
(\ref{susy}) decomposes into four eight-dimensional unit matrices, according
to the decomposition ${\bf 8}_c + {\bf 8}_s + {\bf 8}_c + {\bf 8}_s$
of the thirty-two supercharges, with
coefficients equal to $M$ times $(1+a+b)$, $(1-a-b)$, $(1- a+b)$ and
$(1+a-b)$, respectively. We have BPS multiplets whenever one of these
coefficients vanishes. So we distinguish the following three cases:
\begin{itemize}
\item Choosing 
$a=\pm 1$ and  $b=0$ leads to the $({\bf 8}_v+ {\bf 8}_s)\times({\bf
8}_v+{\bf 8}_c)$ decomposition of the 
$2^8$-dimensional supermultiplet with respect to the rest-frame spin
rotation group SO(8). As always we can
combine multiplets into larger multiplets with higher spin (i.e. by
assigning spin to the Clifford vacuum), but here we concentrate on the
smallest multiplet. Note that this multiplet contains fermions of mixed
chirality. Another characteristic feature is the
presence of a ${\bf 56}_v$ spin representation. This is the multiplet
that comprises the Kaluza-Klein 
states of IIA supergravity compactified on $S^1$, which are the
momentum states of the compactified IIA 
string. Therefore this particular multiplet will be called the KKA
multiplet. 
\item Choosing 
$a=0$ and $b=\pm 1$ leads to the $2^8$-dimensional multiplet $({\bf
8}_v+ {\bf 8}_c)\times({\bf 8}_v+{\bf 8}_c)$. Again we can obtain larger
multiplets of higher spin, but these will not be discussed here. A sign
change in $b$ leads to the 
conjugate multiplet, where ${\bf 8}_s$ and ${\bf 8}_c$ are
interchanged. Obviously the fermions have definite chirality and their
partners in the conjugate supermultiplet carry opposite
chirality. Observe also the absence of ${\bf 56}_v$ states. This
supermultiplet comprises the momentum states of the 
IIB theory and therefore it will be called the KKB multiplet. Clearly, the BPS states associated with a membrane wrapped around
$T^2$ in eleven dimensions constitute KKB multiplets. Observe that
this is crucial for the duality between M-theory and IIB theory, noted
in \cite{JHS}. 
\item The multiplets with 
$\pm a \pm b=\pm 1$ comprise $2^{12}$ states. They do appear in string
theory as mixed states containing both winding and momentum and have a
nonzero oscillator number in order to satisfy the mass-shell
condition. Hence they carry masses of the 
order of the string scale. The smallest multiplet associated with the
lowest spins decomposes as $({\bf 8}_v+ {\bf 8}_c)\times ({\bf
 8}_v+{\bf 8}_c)\times({\bf 8}_v+{\bf 8}_s)$. Again there is a
conjugate multiplet when changing the signs of $a$ and $b$. This class
of BPS supermultiplets will not play a role in what follows. 
\end{itemize}

In the literature one often finds the statement that the IIA and the
IIB theories become indistinguishable when viewed in a
nine-dimensional context, because  the SO(7) decompositions of the IIA
and IIB massless multiplets coincide. Although this is true, it is essential 
to understand that the Kaluza-Klein momentum states for the two theories 
remain different in nine dimensions: for {\it massive} states in nine 
dimensions, 
the rest-frame SO(8) rotation group  coincides with the  SO(8) 
helicity group for massless states in  ten dimensions. 

It is furthermore important that the KKA and KKB multiplets differ 
not only in their spin decomposition, but also carry {\it inequivalent}
charges. We will return to this shortly, but we already note here that a KKA
supermultiplet carries a nonzero SO(2) doublet charge while a  
KKB supermultiplet carries the SO(2) invariant charge. It follows
from the above observations that these charges are mutually exclusive
for these multiplets (but not for the \lq intermediate' multiplets 
with $2^{12}$ states). As explained in the introduction, it is 
of vital importance for duality between the two type-II string 
theories that the winding and the momentum modes of a given 
type-II theory at a given compactification radius constitute 
inequivalent representations and correspond to different kinds of 
string states. 

The above conclusions can also be arrived at by consideration of the 
supersymmetry algebra in eleven dimensions with a membrane charge,
\be
\{Q, \bar Q \} = -i P_{\hat{M}} \G^{\hat{M}} + \ft12 i Z_{\hat{M}\hat{N}}\, 
\G^{\hat{M}\hat{N}}\,, \label{QQZ}
\ee
where we have eleven-dimensional momenta $P_{\hat{M}}$, two-brane charges
$Z_{\hat{M}\hat{N}}$ and 32-component spinor charges. Upon reducing 
this algebra to nine dimensions, assuming that $Z_{\hat{M}\hat{N}}$ 
takes only values in the two extra dimensions labeled by 
$\hat{M}=9, 10$, we obtain for the central-charge matrix $Z^{ij}$, 
\be
Z^{ij} =  Z_{9\,10}\, \d^{ij} - (P_{9} \,\sigma_3 - P_{10} \,
\sigma_1)^{ij}\,. \label{11Dsusyalgebra}  
\ee
From this result, we deduce the general BPS mass formula,
\be
M =  \sqrt{P_9^2 + P_{10}^2}+ | Z_{9\,10}| \, .  \label{BPSmass}
\ee
We will further elaborate on the significance of these formulas later.

\vspace{.5cm}
\noindent{\it  World Sheet Description}\\
The structure of the BPS supermultiplets can also be established
on the basis of the world-sheet superconformal field theory.
In type-II string theory, the two Majorana supercharges can be
represented as contour integrals over world-sheet operators. One
charge, $Q^1_\a$, resides in the left-moving sector and the
other one, $Q^2_\a$, resides in the right-moving sector, so we define
\begin{equation}
Q_\alpha^1= \oint \frac{dz}{2\pi i}\;V_\alpha
(z)\,,\qquad 
Q_\alpha^2= \oint 
\frac{d\bar z}{2\pi i}\;V_\alpha(\bar z)\, .
\end{equation}
In the canonical $q=-1/2$ ghost picture the two covariant left- and
right-moving fermion vertex operators $V_\alpha(z)$ and $V_\alpha(\bar z)$ 
are given  by \cite{FMS} (omitting normal-ordering symbols) 
\bea
V_{\alpha(-1/2)}(z)&=&(\alpha')^{-1/4}\, S_\alpha(z)\, \exp(-\ft12
\phi(z)) \,,\nonumber \\ 
V_{\alpha(-1/2)}(\bar z)&=&(\alpha')^{-1/4}\,S_\alpha(\bar
z)\,\exp(-\ft12\phi(\bar z))\,,  
\label{vertex}
\eea
where $\phi(z)$ ($\phi(\bar z)$) is one of  the left-  (right)-moving
bosonized superconformal ghosts and 
$S_\alpha(z)$, $S_\alpha(\bar z)$ are the spin field vertex operators in 
the ${\bf 16}$ or $\overline{\bf 16}$ chiral spinor representations 
of $SO(9,1)$. Note that~(\ref{vertex}) is valid for both
IIA and IIB string theory, as we refrain from using 
dotted and undotted indices to indicate the chirality.
Whenever this may lead to confusion, the reader should remember to 
simply project onto the corresponding chiral subspaces. 
The ten world-sheet fermions $\psi^M(z)$ can be bosonized in terms 
of five  scalars $\vec \phi$ as
$\exp (i\vec\lambda_{\rm v}\cdot\vec \phi )$, with the SO(1,9) vector 
weights $\vec \lambda_{\rm v} = (0, \ldots, \pm 1, 0,\ldots)$
(thus $\lambda_{\rm v}^2 = 1$). The spin field operators $S_\alpha(z)$ 
can be similarly expressed as $\exp( i\vec\lambda_{\rm s}\cdot\vec\phi(z) )$, 
or $\exp (i\vec\lambda_{\rm c}\cdot\vec\phi(z) )$, where
$\vec\lambda_{\rm s}$ and  $\vec\lambda_{\rm c}$ denote the two 
SO(9,1) chiral spinor weights
$(\pm\frac{1}{2},\pm\frac{1}{2},\pm\frac{1}{2}, 
\pm\frac{1}{2},\pm\frac{1}{2})$, with an even (odd) number of
minus signs for the positive (negative) chirality (and 
$\lambda_{\rm s}^2 =\lambda_{\rm c}^2 = \ft54$). 

In order to compute the supersymmetry algebra from these world sheet 
fields it is convenient to introduce the supercharges  in the 
equivalent $q=+1/2$ superconformal ghost picture. In this ghost
picture $V_\alpha(z)$, for example,  takes the form 
\begin{equation}
V_{\alpha(+1/2)}(z)= (\alpha')^{-3/4} \, \partial X_L^M(z) \,
(\Gamma_M \, S(z))_\a \, \exp(\ft12 \phi(z))\,,
\end{equation}
where the chirality of $S_\a$ is opposite to the chirality of the
$S_\a$ used in the corresponding expression \eqn{vertex}. 
The supersymmetry algebra can now be obtained by computing the
following operator products between the vertex operators in the two
different ghost pictures, 
\bea
V_{\alpha(-1/2)}(z)\, V_{\beta(+1/2)}(w)&\sim&
\frac{1}{z-w}\, {1\over \alpha'} \; (C\, 
\Gamma_M)_{\a\b} \, \partial X_L^M(w) +\cdots \,,
\nonumber \\
V_{\alpha(-1/2)}(\bar z)\, V_{\beta(+1/2)}(\bar w)&\sim&
\frac{1}{\bar z-\bar w}\,{1\over \alpha'} \; (C\, \Gamma_M)_{\a\b}\,
\bar\partial X_R^M(\bar w) +\cdots \,,  
\\[2mm] 
V_{\alpha(-1/2)}(z)\,V_{\beta(+1/2)}(\bar w)&\sim&0\,, \nonumber
\label{ope}
\eea
where $C$ is the ten-dimensional charge-conjugation matrix. Taking the
contour integrals and converting to the nine-dimensional gamma indices
introduced earlier, we find the $\{Q^i,\bar Q^j\}$ anticommutator in terms
of the nine-dimensional momenta and the right- and left-moving
zero-mode momenta,  
\bea
p_L & = & {1\over \alpha'}\, \oint\frac{dz}{2\pi i}\; 
i\,\partial X^9_L(z)  =
\frac1{\sqrt{\alpha'}} \, \biggl(\frac{m}{T} - nT \biggr)\,,
\nonumber \\ 
p_R & = &  {1\over \alpha'}\,\oint\frac{d\bar z}{2\pi i} \;
i \,\bar\partial X^9_R(\bar z) = 
\frac1{\sqrt{\alpha'}}\, \biggl(\frac{m}{T} + nT\biggr) \,,
\label{plr}  
\eea
where we measure the compactification radius $R$ in string units 
by means of a dimensionless parameter $T=R/\sqrt{\alpha'}$. The
integers $m$ and $n$ denote the momentum and winding numbers,
respectively.  This yields the following 
supersymmetry algebra for the IIA/B superstrings in nine space-time 
dimensions
\bea
\{ Q^1,\bar Q^1\}&=& -i P_\mu\,\gamma^\mu  - p_L \,\Gamma^{11}\,,\nonumber \\
\{ Q^2,\bar Q^2\}&=& -i P_\mu\,\gamma^\mu - p_R\, \Gamma^{11}\,, \nonumber\\
\{ Q^1,\bar Q^2\}&=& 0\, . 
\label{wssusy}
\eea
Comparing with the previously derived supersymmetry algebra~(\ref{susy}),  
it is now obvious that the central charges 
are just linear combinations of the internal left- and right
momenta $p_L$ and $p_R$. To be more precise, in the IIA and 
the IIB theory the central charge matrix $ Z^{ij}$ takes one of
the two alternative forms (up to an overall sign),
\be
 Z^{ij} = \left\{ \begin{array}{ll} \ft12( p_L + p_R)\,\d^{ij}  +
\ft12( p_L - p_R) \,(\sigma_3)^{ij} &\mbox{ for IIB }\\[1mm]
\ft12( p_L - p_R)\,\d^{ij}  + \ft12( p_L +
p_R)\, (\sigma_3)^{ij} &\mbox{ for IIA } \end{array} \right.
\label{wsZ}
\ee
This proves our assertion that the momentum and winding BPS
states constitute inequivalent supermultiplets. The IIA
momentum  states and the IIB winding states are in the KKA
representation, whereas the IIA winding states and the IIB momentum
states are in the KKB representation. This ensures that the two
decompactification limits $T\to 0$ and  $T\to \infty$ lead to different
theories. Moreover, it proves that type-II 
string compactifications on circles of different radii must be
inequivalent. And finally, it is clear that no symmetry enhancement
will take place when the momentum and the winding states have
coinciding masses, as these states are always
distinctly different. This is in accord with the fact that  no
gauge symmetry enhancement is possible in the conformal field theory. 

The emergence of different representations for the momentum and
winding states can also be understood in terms of the corresponding 
covariant physical vertex operators. To write them down for the 
compactified theory in nine dimensions, we again make use of the 
SO(1,9) covariant ghost and spin field vertex operators. The vertex
operators for the Kaluza-Klein 
and winding states in nine dimensions can be directly obtained from 
the vertex operators of the massless states in ten dimensions by 
splitting the physical momenta as in~(\ref{wssusy}). More precisely,
we consider the Ramond-Ramond operators 
\bea
&& \exp \big(ip_\mu X_L^\mu (z) + ip_L X_L^9(z)\big)\; 
    \bar u_L^{\alpha}(p) S_{\alpha}(z) \,
   \exp (- \ft12 \phi (z) )        \nonumber \\
&\times & \exp \big(ip_\mu X_R^\mu (\bar z) + ip_R X_R^9(\bar z)\big)\; 
    \bar u_R^{\beta}(p) S_{\beta}(\bar z)\, 
    \exp ( - \ft12 \phi (\bar z) ) \,,
\label{operators}
\eea
where the 16-component spinors $u_L(p)$ and $u_R(p)$ denote the chiral  
SO(1,9) spinor polarizations of the left- and right-moving
states (so that we have implemented a GSO projection) and
the $p_\m$ are the values taken by the nine-dimensional 
momentum operators $P_\m$. Again we refrain from using dotted 
and undotted SO(1,9) spinor indices and we leave the chirality of the
spin fields and therefore of the polarization spinors
unspecified. Note, however, that the chirality of the $S_\a$ must be
the same as in \eqn{vertex}. We recall also that the ghost and
spinor weights for the vertex operator 
must be chosen in accordance with the locality requirement
(see \cite{KLLSW} for a detailed discussion of this point).

To the operators \eqn{operators} we must apply the physical state
condition which follows from requiring that they commute with the
left- and right-moving BRST operators. The relevant part of 
the left-moving such 
operator is proportional to $\pa X^M_L(z)\,\psi_{L\,M}(z)\,
\exp(\phi(z))$; the formula for its
right-moving counterpart is similar. In this way we recover first of 
all the mass-shell condition  
$- p_\mu p^\mu = p_L^2 = p_R^2$, where the last equality is valid only
for states without oscillator excitations. Secondly, 
we obtain the Dirac equation for the spinor polarizations. Written
with nine-dimensional gamma matrices, this yields, 
\be
( ip_\mu \gamma^\mu + p_L \Gamma^{11} )
  u_L(p)= ( ip_\mu \gamma^\mu + p_R \Gamma^{11})
  u_R(p) = 0 \,.
\label{Dirac}
\ee
These conditions reduce the number of physical spinor polarizations 
from 16 to 8, so that the vertex operators \eqn{operators} describe 
$8\times 8 =64$ states for given momentum. When combined with the 
Neveu-Schwarz sector these states comprise full BPS supermultiplets. 
In obtaining the SO(8) representations in accord with our earlier  analysis,  it is important to realize that the chirality of the 
polarization spinors is opposite to that of the corresponding $S_\a$. 

The above results are in precise correspondence with our previous
analysis of the superalgebra relations \eqn{wssusy}. The mass-shell
condition tells us that $p_L = \pm p_R$, and depending on this sign,
we get either the same or different SO(8) representations from the  
physical state condition~(\ref{Dirac}). Thus winding and momentum
states indeed constitute inequivalent supermultiplets. 

Finally let us discuss the choice of the chirality for the spinors 
in \eqn{vertex} and in \eqn{operators}. Clearly we only need to
distinguish between equal and opposite 
chirality for the left- and right-moving spinor fields. On the other
hand, switching the relative chirality, e.g. by changing the
chirality of $S_\a(z)$, and correspondingly of $u_L$, can be
compensated for by assigning an opposite 
momentum  $p_L$ to that state, leaving $p_R$ unchanged. This
corresponds to interchanging the winding and the momentum numbers $m$ and
$n$ in \eqn{plr}, together with the interchange of  $T$ with $1/T$. So the
states and the corresponding supermultiplets remain the same;  what
changes is only the notion of a momentum and a winding state. Clearly,
there is a type-IIA and a type-IIB description, but of a single theory. In the
decompactification limits $T\to 0$ and $T\to \infty$, one is left with
inequivalent supermultiplets as the mass of one supermultiplet vanishes and that of its inequivalent
counterpart is pushed to infinity. 

It is equally straightforward to analyze the \lq intermediate' BPS multiplets
with $2^{12}$ states from this point of view.  However, the corresponding
vertex operators are more complicated due to oscillator
contributions, which  modifies the relation between $p_L$ and 
$p_R$.

\vspace{0.5cm}

\noindent{\it  $N=2$ Supergravity in Nine Dimensions}\\
\setcounter{table}{0}
We will now use $N=2$ supergravity in $D=9$ dimensions together with
some basic input from string
theory to obtain independent confirmation of the result that
the momentum and winding states are in different supermultiplets. 
Let us first discuss some features of the massless fields 
which constitute $N=2$  supergravity in  nine
dimensions. This theory has already been  
discussed in the literature; in particular, its relation 
to string theory and IIA/B duality was studied in
\cite{Bergshoeff:1995as}, so we will 
be brief here. The focus of our attention is the  coupling of the massless theory to the massive BPS states
that we discussed above. In particular, we want to exhibit the
coupling of the nine-dimensional gauge fields to the BPS states. 

In nine dimensions there is only one $N=2$ supergravity theory, 
whose scalar sector is governed by an SL$(2,\bR)/{\rm SO}(2)$
non-linear $\sigma$-model, and which therefore exhibits an invariance 
under a nonlinearly realized SL$(2,\bR )$. In addition there is an
invariance under SO(1,1), which can be systematically understood from 
combining ordinary dimensional analysis with scale 
transformations on the compactified coordinate~\cite{cargese}. From the IIB supergravity  perspective, the SL$(2,\bR)$ 
originates from the SL$(2,\bR )/{\rm SO}(2)$ coset 
structure and the SL$(2,\bR)$ symmetry which are already present
in ten dimensions~\cite{Schwarz:1983qr}. From the perspective of eleven-dimensional 
supergravity \cite{CJS}, on the other hand, these are just the \lq hidden'
symmetries obtained by reducing the theory from eleven to nine 
dimensions on the torus $T^2$. In this reduction, the diffeomorphism 
symmetry in the compactified dimensions is \lq frozen' to a rigid 
GL$(2,\bR) = {\rm SL}(2,\bR) \times {\rm SO}(1,1)$ symmetry.
Similarly, the full Lorentz symmetry in eleven dimensions is  
reduced to SO$(1,8)\times {\rm SO}(2) \subset {\rm SO}(1,10)$, where
SO(2) is converted into the R-symmetry corresponding to the automorphism 
group of the nine-dimensional $N=2$ superalgebra. 

Identifying the various transformations, one readily obtains the 
various quantum numbers,
without the need for a detailed dimensional reduction. 
We denote the bosonic fields of eleven-dimensional
supergravity by $\hG_{\hat{M} \hat{N}}$ and 
$\hA_{\hat{M} \hat{N} \hat{P}}$. The bosonic fields of IIA supergravity 
are denoted by $G_{MN}$, $C_{M}$, $C_{MN}$, $C_{MNP}$ and $\phi$, and
those of IIB supergravity by $G_{MN}$, $A_{MN}^{\,\a}$, $\phi^\a$ and
$A_{MNPQ}$. Here the index $\a$ is associated with SL$(2,\bR)$. The fields
of $N=2$ nine-dimensional supergravity are the metric 
$g_{\m\n}$, three scalars $\sigma$ and $\phi^\a$, three abelian gauge
fields $B_\m$ and $A_\m^{\,\a}$, two antisymmetric tensors
$A_{\m\n}^{\,\a}$ and a three-rank antisymmetric gauge field
$A_{\m\n\rho}$. The fields and their SO(1,1) weights are summarized in
the table. We use the Einstein frame, so that the metric is
invariant under SO(1,1). The scalar fields $\phi^\a$ characterize the
coset representative of SL$(2,\bR)/{\rm SO}(2)$. They satisfy a
constraint $\phi^\a \phi_\a =1$ and are subject to local SO(2)
transformations, so that they correspond to one complex field. The
scalar $\exp(\sigma )$ will be defined as $G_{99}$, the IIB metric in the
compactified dimension. The determinant of the eleven-dimensional
metric in the two compactified directions is then equal to 
$\exp( -\ft43\sigma )$. We have ignored certain nonlinear features of the 
relationship with the higher-dimensional fields. On the other hand,
the assignments are also relevant for the massive  
Kaluza-Klein states in the $T^2$ and $S^1$
compactifications~\cite{cargese}. 

\begin{table}

\begin{center}
\begin{tabular}{||c|c|c|c|c||}
\hline $D=11$ & IIA & $D=9$ &  IIB & SO(1,1) \\ 
\hline \hline 
$\hG_{\mu \nu}$ & $G_{\m \n}$  & $g_{\m \n}$  & $G_{\m \n}$ & $0$ \\
\hline 
$\hat{A}_{\mu {\scriptscriptstyle \,9\,10}}$  & $C_{\mu
{\scriptscriptstyle\,9}}$ &  $B_{\m}$ & $G_{\m {\scriptscriptstyle\,9}}$
& $-4$\\ 
\hline 
$\hG_{\mu {\scriptscriptstyle\,9}}$, $\hG_{\mu
{\scriptscriptstyle\,10}}$  & $G_{\mu {\scriptscriptstyle\,9}}$ ,
$C_{\mu}$  & $A_{\mu}^{\, \alpha}$ & $A_{\m
{\scriptscriptstyle\,9}}^{\, \alpha}$ & $3$  \\  
\hline 
$\hat{A}_{\m \n {\scriptscriptstyle\,9}}$, $\hat{A}_{\m \n
{\scriptscriptstyle\,10}}$ & $C_{\m \n {\scriptscriptstyle\,9}}, C_{\m
\n}$ &  $A_{\m\n}^{\,\a}$ &  $A_{\m\n}^{\,\a}$  & $-1$  \\ 
\hline 
$\hA_{\m \n\rho}$  & $C_{\m \n \rho}$  & $A_{\m\n\rho}$  &
$A_{\m\n\rho\sigma}$ & 2  \\ 
\hline 
$\hat G_{{\scriptscriptstyle 9 \,10}}$, $\hat{G}_{{\scriptscriptstyle9\,
9}}$, $\hat G_{{\scriptscriptstyle 10\,  10}}$ & $\phi$,  
$G_{{\scriptscriptstyle 9\,9}}$, $C_{\scriptscriptstyle  9}$ &
$\left\{ \begin{array}{l} \phi^\a  \\ \exp (\sigma ) \end{array}
\right. $ & $\begin{array}{l} \phi^\a \\ G_{{\scriptscriptstyle
9\,9}} 
\end{array}$ & $\begin{array}{r} 0 \\  7 \end{array}$
\\ 
\hline 
\end{tabular}
\vspace{.3cm}
\caption{The bosonic fields of the eleven dimensional, type-IIA,  
nine-dimensional $N=2$ and type-IIB  supergravity theories. 
The eleven-dimensional and ten-dimensional indices, respectively, are 
split as $\hat{M} =(\mu, 9,10)$ and $M=(\mu,9)$, where $\mu = 0,1,\ldots 8$. 
The last column lists the SO(1,1) scaling weights of the fields.}
\end{center}

\end{table}

Now we consider the three abelian vector gauge fields in the
nine-dimensional theory, which decompose into a singlet and a
doublet under SL$(2,\bR)$. Note that their origin is rather different when
viewed from the IIA  and from the IIB side. The singlet field
is the graviphoton from the IIB side, so it must couple to the IIB
momentum states. The doublet fields originate from the IIB doublet of
tensor fields, so they couple to the IIB winding states. It thus follows that 
the IIB momentum states constitute KKB
states (by definition) whereas the IIB winding states constitute KKA
multiplets. The second KKA charge can only be understood beyond string
perturbation theory; the degeneracy in the winding states is due to
winding of fundamental and D-strings. 

The pattern is the same, but complementary on the IIA side. Here the
momentum states carry the doublet charges, so they constitute (again by
definition) KKA multiplets. Accordingly, the two graviphotons originating
from eleven dimensions transform as an SL$(2,\bR)$ doublet.
The degeneracy in the momentum states can thus be understood from 
eleven-dimensional supergravity, as the doublet charges 
find their origin in the $T^2$ on which the theory is compactified. 
The winding states couple to the singlet field, which originates 
from the IIA tensor field. Hence the IIA winding states constitute 
KKB multiplets. Alternatively these states can be understood as
membranes wrapped around the M-theory torus \cite{JHS}, because, as
we have shown before, these constitute the same supermultiplets.

\newpage
\noindent{\it  Coupling to BPS Supermultiplets}\\
One may contemplate the construction of a nine-dimensional field
theory consisting of $N=2$ supergravity coupled to an infinite tower
of BPS supermultiplets with a two-dimensional charge lattice
$(q_1,q_2)$ for the KKA states and a one-dimensional lattice of
charges $p$ for the KKB states. This theory encompasses 
both eleven-dimensional supergravity (compactified on $T^2$) and IIB
supergravity (compactified on $S^1$). The usual T-duality is trivial
for this theory. It is not associated with any symmetry and only
amounts to certain field redefinitions. We know that the theory is free 
from inconsistencies in each of these sectors separately and it is an
interesting question whether such a \lq dichotomic' field theory could be
(classically) consistent to all orders. In low orders of perturbation
theory, its 
short-distance behaviour should be relatively mild as it can be viewed
as a combination of known supergravity theories. Of course, this is
not truly an effective field theory as the masses of the various
states will never be light simultaneously with respect to the string
scale. The theory is manifestly invariant under SO$(1,1)$ and under
SL$(2,\bZ)$. The latter is the integer-valued subgroup of SL$(2,\bR)$
that leaves the charge lattice of the KKA states invariant. There is a
formulation in which the SL$(2,\bR)$ is linearly realized, also in the
presence of the BPS states. In that case the massive fields transform
only under the local (composite) SO(2) and not directly under
SL$(2,\bR)$. However, the KKA  fields have a minimal coupling with 
respect to $q_\a\,A^\a_\m$, which, in order to remain invariant under 
the integer-valued subgroup, requires the charges to transform 
covariantly under this subgroup. The KKB fields have a minimal 
coupling to $p\, B_\m$, which is SL$(2,\bR)$ invariant. 

It should be clear that the theory will exhibit ten- or 
eleven-dimensional Lorentz invariance only in certain limits.
For the KKA states with charges $q_\a$, and KKB states with
charge $p$, respectively, the BPS mass formula in the 
nine-dimensional Einstein frame is given by
\be
M = m_{\scriptscriptstyle \rm KKA} \,\vert q_\a\,\phi^\a\vert  
 +  m_{\scriptscriptstyle \rm KKB} \, |p| \,  ,  
 \label{BPS2}
\ee
where $m_{\scriptscriptstyle \rm KKA}$ and $m_{\scriptscriptstyle\rm KKB}$ 
denote two different mass scales, whose product is inversely 
proportional to $\alpha'$. Here we made use of the fact that the mass 
should be SL$(2,\bZ)$ and SO(1,1) invariant in the Einstein frame. 

As noted in \cite{JHS} the mass formula \eqn{BPS2} is entirely consistent
with that of a membrane wrapped around a torus with modular parameter
$\tau\equiv \tau_1+i\tau_2$. Here we should point out that the
supersymmetry algebra for a fundamental supermembrane gives 
rise to precisely the algebra \eqn{QQZ} with 
$Z_{\hat{M}\hat{N}}$ describing the winding of the membrane over 
some compact space \cite{DWPP}. In the case of a torus with area $A$,
the BPS mass formula follows directly from \eqn{BPSmass} and reads 
(in eleven-dimensional Planck units)
\bea
M =  {1\over {\sqrt{A\, \tau_2}}}\, \vert q_1 -\tau\,q_2\vert + 
A T_{\rm m}\,| p|  \,, \label{membrane-bps}
\eea
where $T_{\rm m}$ denotes the  supermembrane tension, $q_{1,2}$ label
the momentum modes on the torus and $p$ is the number of times the
membrane is wrapped (including orientation) over the torus. This
formula agrees with the one previously derived in \cite{RT} on 
the basis of a semi-classical approximation. We refrain from
indicating how the modular parameter is related to the fields
$\phi^\a$ but simply note that both formulae are invariant under
SL$(2,\bZ)$.

The  formula~(\ref{BPS2}) can now be interpreted in two different ways. From the 
perspective of IIA string theory, one of the $q_\a$ is the
IIA Kaluza-Klein momentum number, while the other is the D0 charge;
as is well known, the mass of the D0 branes is inversely proportional 
to the IIA string coupling constant~\cite{Joe}. Then $p$ is the 
IIA winding number. Conversely, from the IIB perspective, $q_1$ and 
$q_2$ are the winding numbers of the elementary string and of the solitonic 
D1 string (which corresponds to a D0 brane in the IIA description). Now the 
SL$(2,\bZ)$ is a strong-weak coupling duality, 
as it interchanges the elementary strings with the D1 strings. 
The modular parameter associated with the fields $\phi^\a$ is the IIB dilaton
which contains the IIB string coupling constant. From this perspective
the integral charge lattice follows from a Dirac-type quantization
condition. The integer $p$ is just the IIB Kaluza-Klein momentum number. 

The question that remains is, of course, what IIA/B duality can
teach us about M-theory and its fundamental degrees of freedom.
The theory we referred to as \lq dichotomic' above transcends both 
eleven-dimensional supergravity and IIB supergravity. The above results can 
be interpreted as evidence that 
M-theory is just the fundamental supermembrane. Supermembrane theory
may not suffer from the incompleteness of perturbative string
theory. Unlike superstring theory, which has both a string tension 
as well as a coupling constant, it has no conventional perturbative 
expansion as its only parameter is the membrane tension $T_{\rm m}$. 
As is evident from \eqn{11Dsusyalgebra} and \eqn{membrane-bps},
both the Kaluza-Klein doublet states and the winding states arise 
naturally upon compactification to nine dimensions. Likewise, the 
perturbative massive string states, which have no analog in the 
(uncompactified)
supermembrane, can emerge out of the continuous supermembrane
spectrum \cite{dWLN} in the reduction from eleven to ten
dimensions (recall that the excited  superstring states cannot be
combined into massive $D=11$ multiplets). This indicates that  the quantum
supermembrane is not only  a second quantized, but also a 
non-perturbative theory from the very outset --- like M-theory.

\newpage
\noindent{\it Acknowledgements}\\[1mm]
We would like to thank Michael Green, Joseph Polchinski,  Jorge Russo, Bert
Schellekens and Nathan Seiberg for discussions. B.d.W.\ is grateful to
the Alexander von Humboldt-Stiftung for supporting his stay at the AEI
as part of the Humboldt Award Program. This work is supported in part by
the European Commission TMR Program under contract  ERBFMRX-CT96-0045, in which
the Humboldt University at Berlin and Utrecht University are associated.

  
\end{document}